# Beyond the Hype: Critical Analysis of Student Motivations and Ethical Boundaries in Educational AI Use in Higher Education

**Adeleh Mazaherian[1,*], Erfan Nourbakhsh[2]**

[1]*Department of Education and Psychology, University of Isfahan, Isfahan, Iran, adelehmazaherian@gmail.com*

[2]*Computer Engineering Department, University of Isfahan, Isfahan, Iran, erfannourbakhsh2001@gmail.com*

## Abstract.

The rapid integration of generative artificial intelligence (AI) in higher education since 2023 has outpaced institutional preparedness, creating a persistent gap between student practices and established ethical standards. This paper draws on mixed-method surveys and a focused literature review to examine student motivations, ethical dilemmas, gendered responses, and institutional readiness for AI adoption. We find that 92% of students use AI tools primarily to save time and improve work quality, yet only 36% receive formal guidance, producing a de facto "shadow pedagogy" of unguided workflows. Notably, 18% of students claimed to integrate AI-constructed material into assignments, which suggests confusion about integrity expectations and compromises the integrity of the assessment. Female students were more worried about abuse and distortion of information than male students, revealing a gendered difference in awareness of risk and AI literacies. Correspondingly, 72% of educators use AI, but just 14% feel at ease doing so, and this signifies restricted training and uneven policy responses. We argue that institutions must adopt comprehensive AI literacy programs that integrate technical skills and ethical reasoning, alongside clear AI-use policies and assessment practices that promote transparency. The paper proposes an Ethical AI Integration Model centred on literacy, gender-inclusive support, and assessment redesign to guide responsible adoption, protect academic integrity, and foster equitable educational outcomes in an AI-driven landscape.

## 1. Introduction

The years since 2023 represent a landmark period in higher education technology, directly as a result of the widespread and blockbuster uptake of generative artificial intelligence (AI) tools such as ChatGPT. This has disrupted the workflows of higher education students' research, writing, studying, and content interaction (Walter, 2024). Survey data estimates that by early 2025, student uptake of generative AI tools stands at 92% (up from 66% in 2024), with 88% deploying AI for academic assessments (Freeman, 2025). Other international survey data support the findings, with an estimated 86% of students already using AI in their education (Digital Education Council, 2024; Campus Technology, 2024). This bottom-up emergence of AI/machine learning usage has led to a 'shadow pedagogy', where students construct AI-assisted workflows outside of the formally designated institutional teaching practices (Walter, 2024). The magnitude of this disruption means that institutions must grapple with pivotal issues of the learning and assessment paradigm and the implications of academic integrity in the digital era.

The relationship between student behaviour and institutional preparedness is negative in higher education despite the prevalence of use. Almost all students use AI applications, but only 36% have been formally trained on how to use them ethically, and less than half of instructors believe they are prepared to teach students how to integrate AI responsibly (Freeman, 2025; Zawacki-Richter et al., 2024). This is a concerning statistic considering the ethically murky terrain that AI programs present. For instance, almost one-in-five students admit to copying and pasting outputs from AI programs, word-for-word, into their submitted work (Freeman, 2025). Gender also emerges as an important variable in institutional preparedness and use. Specifically, female students are more likely than male students to be concerned about the potential for academic dishonesty and the dissemination of misinformation (Freeman, 2025; Russo et al., 2025).

This analysis addresses four research questions examining AI integration challenges in higher education through early 2025:

- **RQ1**: What is the motivation for AI adoption by students in the higher education context? It is important for the institutions to know the motivation. Data reveals that the most prominent motivators are pragmatic, with 51% of responses seeking to use AI solutions to save time and 50% to increase quality (Freeman, 2025).
- **RQ2**: What is the level of difference between student activity and the perceived ethical limits? This difference is at odds with the conventional notions of academic integrity, as students undertake activities that they frequently regard as unethical (Freeman, 2025).
- **RQ3**: What is the impact of demographic factors, especially gender, on the AI use differences? Gender discrepancies in usage and risk tendencies require AI acquisition strategies to be inclusive (Freeman, 2025).
- **RQ4**: How and what might these findings mean for higher education institutions and policy? The situation today regarding unguided AI usage potentially causes issues for higher education institutions when it comes to academic integrity and educational quality (U.S. Department of Education, 2025).

Whereas students have accepted AI due to improved efficiencies and performance augmentation, ethical gaps and gendered responses portray complications necessitating prompt institutional intervention. This paper looks into AI adoption in tertiary education using recent surveys and policy frameworks (Freeman, 2025; UNESCO, 2023). It analyzes student motivations, ethical gaps, gender inequalities, and institution-readiness, and recommends comprehensive, human-oriented AI integration frameworks to cover technology and ethical aspects.

## 2. Background

The integration of generative artificial intelligence (AI) into higher education since 2023 has transformed academic practices, raising critical questions about pedagogy, ethics, and equity (Walter, 2024; Freeman, 2025). This section explores the evolution of AI technologies, their ethical implications, and gender-based adoption patterns, drawing on systematic reviews and policy frameworks to contextualize the rapid shift toward student-driven AI use (UNESCO, 2023; Ng et al., 2024).

### 2.1 Evolution of AI in Higher Education

AI systems used in the early stages of education were not generalized to various subjects (Walter, 2024; Zawacki-Richter et al., 2024). Specifically, ITS (intelligent tutoring systems) were used for learner support for student courses like mathematics and language learning. However, this system was used to follow rule-based adaptive learning technologies. AI was a useful tool for supporting individualized learners by providing various means of feedback and remediation to elementary human tutors. However, these systems could not create an open-ended human process. This would define the evolution towards generalized AI.

The large language models and generative AI that appeared in 2022, such as ChatGPT, represent a quantum leap in technology that can accompany a wide range of educational activities, from essay writing and text comprehension to drafting research ideas (Ng et al., 2024; Tillmanns et al., 2025). In higher education, AI democratized access to the latest technological capabilities as students begin to apply AI to perform tasks that previously required intellectual labor, professional knowledge, or access to substantial resources (Bond et al., 2024). In this regard, we observe the reversal of the technology adoption model: the traditional genesis from institutional bodies and teachers who determine the policy of integration and use of technologies in the curriculum has changed to a model characterized by bottom-up and informal initiation and appropriation of technology by students (Zawacki-Richter et al., 2024). Regarding policies, the recommendations of UNESCO in 2023 on the use of generative AI and the principles of responsible implementation of the U.S. Department of Education (2025) point out humanistic goals, data privacy, and transparency, although we observe an uneven impact in their implementation (UNESCO, 2023; U.S. Department of Education, 2025). The systematic reviews published in the last months state that the application of AI in curricula and assessments is in development, where the benefits of personalized learning contend with the risk of over-reliance (Liang et al., 2025).

### 2.2 Ethical Considerations

AI ethical issues in higher education go beyond technical competency to basic issues of academic integrity, fairness, and human agency (Zheng et al., 2025; García-Peñalvo, 2025). Conventional models of academic misconduct, centered on plagiarism from human sources or unauthorized collaboration, are insufficient for generative AI, which may create original material indistinguishable from student submissions (Ng et al., 2024). These create ambiguities among students and educators about proper assistance (e.g., grammar checking) and misconduct (e.g., complete content generation) (Freeman, 2025).

New ethical frameworks call for nuanced responses, such as framing "appropriate use" according to assignment type, learning outcomes, and levels of dependence on AI (Chen et al., 2024; Zheng et al., 2025). For example, AI literacy skills now encompass critical evaluation of outputs, awareness of biases, and ethical judgment, as set out in UNESCO's 2024 student framework (UNESCO, 2024). However, dangers remain, such as the reinforcement of inequalities by algorithmic bias and data privacy issues in educational AI applications (Wargo & Anderson, 2024). Institutions have to offset AI advantages of greater efficiency and access against protections against excessive dependence, which may damage critical questioning and original research (Holmes et al., 2019). Systematic analyses cite 53 cases of bias dangers and 46 of privacy issues, and highlight the need for ethical guidance (Zheng et al., 2025).

### 2.3 Gender and AI Adoption

Gender disparities in technology adoption have long influenced educational outcomes, and AI integration in higher education reveals similar patterns, with male students often exhibiting higher confidence and enthusiasm while female students express greater caution and ethical concerns (Freeman, 2025; Russo et al., 2025). Historical research shows persistent gaps in self-efficacy and risk perception, where female students prioritize potential negative consequences like misinformation or academic penalties over benefits (Chen et al., 2024).

In AI in higher education settings, these inequalities appear in use motivations and attitudes; male students are more likely to utilize AI in skill-development (36% compared to 22% females), and females have greater misconduct anxiety (Freeman, 2025). Females have greater AI anxiety and lower favourable attitudes, creating adoption disparities of 10-40% (Russo et al., 2025; Koning et al., 2025). This "gender digital gap" has the potential to widen inequities in AI competency development, potentially inhibiting females' contributions to AI fields (UNESCO, 2024). Redress involves inclusive courses building on diverse strengths, such as females' ethical thinking, and developing confidence in generative, non-competitive learning settings (Ng et al., 2024). Reviews indicate females view AI as less socially favourable, adding to concerns about job loss and communication (Russo et al., 2025).

## 3. Methodology

This study synthesizes recent data and literature to examine AI adoption in higher education, focusing on student motivations, ethical gaps, gender disparities, and institutional readiness through early 2025 (Freeman, 2025; Zawacki-Richter et al., 2024). By combining quantitative survey data with qualitative policy and literature analyses, the methodology ensures a comprehensive approach to addressing the research questions, grounded in established frameworks for technology adoption and ethics (Chen et al., 2024; UNESCO, 2023).

### 3.1 Data Sources

This analysis takes a synthesis of information from a secondary sources approach to exploring AI adoption from the higher education market, largely from the 2025 HEPI/Kortext Student Generative AI Survey, which contains quantitative information regarding student behaviour, motives, and attitudes (Freeman, 2025). Surveying 1,041 university undergraduates who are full-time university students, this provides important statistics pertaining to adoption rates, ethical behavior, and gender inequity. Teachers' perceptions complete these in teaching with AI and training gaps reports globally (Zawacki-Richter et al., 2024; Microsoft Corporation, 2025). UNESCO, U.S. Department of Education, and EDUCAUSE papers on policy provide patterns and frameworks (UNESCO, 2023; U.S. Department of Education, 2025; EDUCAUSE, 2024). Systematic literature reviews of AI literacy, AI ethical behaviour, and teacher development complete the analysis (Liang et al., 2025; Nguyen et al., 2024; Zheng et al., 2025; Bond et al., 2024; Tillmanns et al., 2025; Holmes et al., 2019). Surveys of the higher education market are complemented by Ellucian (2024), Ithaka S+R (Cooper & Ruediger, 2025), and Campus Technology (2024). Strengths are outweighed by the questionnaire data dependence, the potential response bias, and only the 2023–2025 time period, such that whole patterns are missed. Neglecting everything other than the higher education market avoids an excessive level of overgeneralization.

### 3.2 Analytical Approach

Mixed-methods analytical design involves quantitative descriptive statistics of surveys (e.g., adoption rates, gender comparisons) combined with thematic analysis of literature reviews and policy documents (Freeman, 2025; Zawacki-Richter et al., 2024). Quantitative data are examined for trends and disparities by cross-tabulations, e.g., student behaviour versus ethical perceptions. Qualitative synthesis extracts thematic patterns such as ethical gaps and institutional preparedness, underpinned by technology adoption and gender and education theoretical frameworks (Chen et al., 2024; Russo et al., 2025). Triangulation across sources allows greater validity, though understanding biases such as self-reporting. The approach responds to the research questions through structured, evidence-driven mapping of motivations (RQ1), ethical gaps (RQ2), gender influencing factors (RQ3), and implications (RQ4).

## 4. Findings

The findings illuminate the rapid and complex integration of AI in higher education, revealing high student adoption, significant ethical challenges, gender-based differences, and lagging institutional preparedness (Freeman, 2025; Zheng et al., 2025). Supported by survey data and visual representations (Figures 1–3), this section addresses the four research questions, detailing motivations, ethical gaps, gender disparities, and educator readiness to provide a comprehensive picture of AI's impact (Walter, 2024; UNESCO, 2023).

### 4.1 Student Motivations for AI Use

Adoption of generative AI tools in universities occurs largely due to pragmatic reasons, with 92% of students adopting AI by early 2025 compared to 66% in 2024 (Freeman, 2025). Survey results indicate 51% of students name time-saving as a top reason, such as using AI to write essays or summarize research, with 50% looking for quality enhancements, such as improved clarity or coherence of written work (Freeman, 2025). Figure 1 demonstrates AI tool adoption breadth, with 64% of students utilizing AI to generate text, 39% to improve writing, and 28% to brainstorm or generate ideas. Secondary reasons are skills development (28%) and interest in AI potential (15%), suggesting an array of practical and curiosity-driven motivators (Freeman, 2025). Subsequent surveys endorse widespread use, with 86% of students using AI, 69% for searching information, and 42% checking grammar (Campus Technology, 2024).

Students utilize multiple tools in advanced workflows, such as pairing text generation with editing or analysis tools, signifying a student-led approach to exploiting AI-generated benefits to enhance academic agility (Walter, 2024). But even this student-led approach happens largely outside institution-provided frameworks, creating a "shadow pedagogy" of students creating AI practices but without formal direction (Walter, 2024).

*Figure 1: Usage patterns of generative AI tools among higher education students*

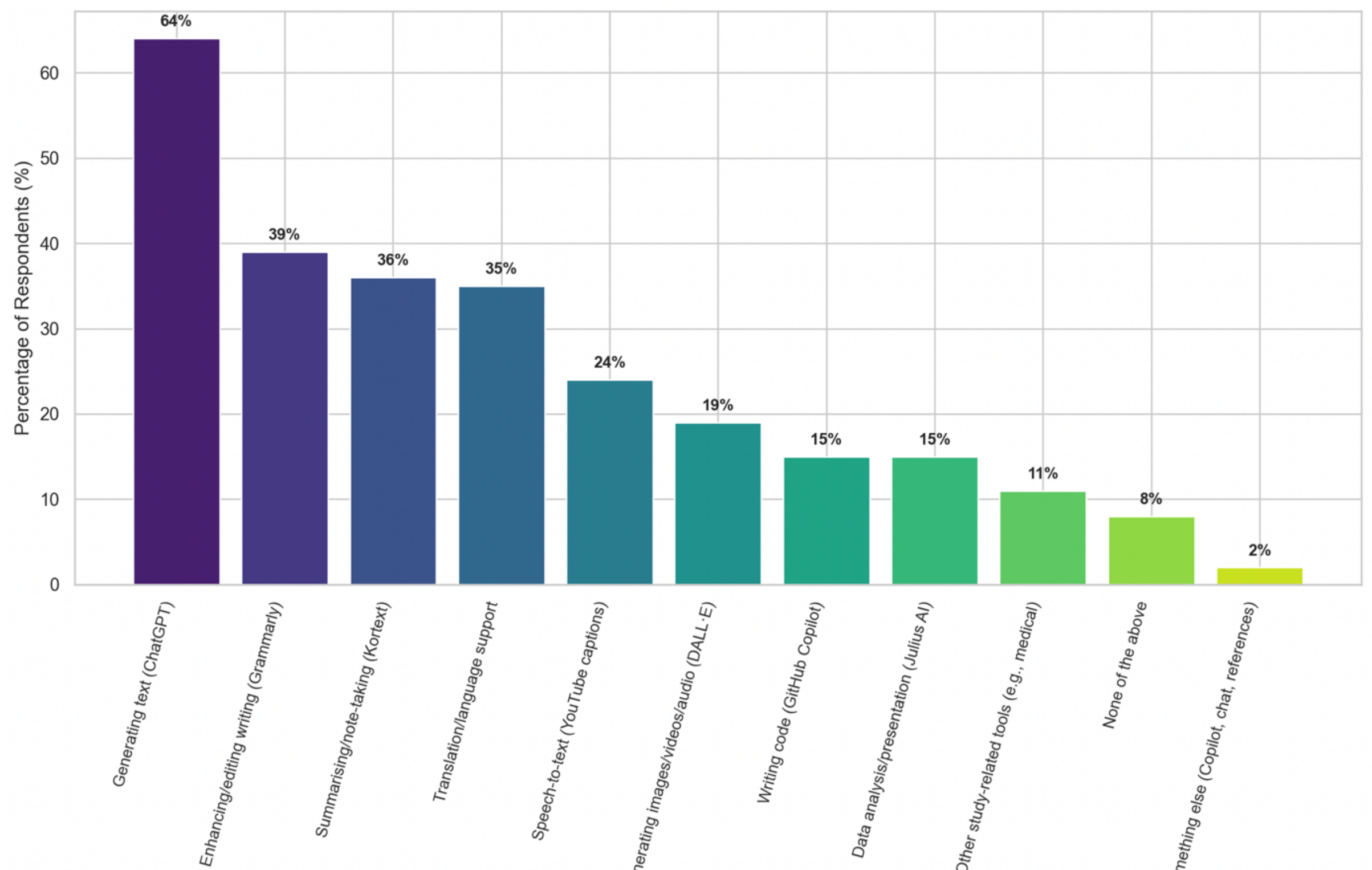

*Source: Freeman (2025)*

### 4.2 Ethical Gaps in AI Use

There is a gap between student AI use and knowledge of ethical limits, defying traditional concepts of academic integrity. Though 88% of students use AI in assessment, only 36% have institutionally provided guidance on proper use, and 18% of students confessed to pasting AI-generated work into submissions (Freeman, 2025). Figure 2 illustrates this disconnection, where only 20% of students find copying AI output acceptable, but more students practice this due to confusion or rationalization (Freeman, 2025). Students tend to rationalize such behaviour by perceiving some assignments as ``busywork" or finding AI usage to be untraceable, an attitude brewed by uneven institution-wide policies (Ng et al., 2024). Privacy threats and algorithmic biases further blur ethical concerns because students unwittingly leak sensitive information or base submissions on biased results (Zheng et al., 2025). Lack of definable guidelines encourages permissive culture testing of ethical limits and defies academic integrity (García-Peñalvo, 2025).

*Figure 2: Comparison of student AI use practices versus perceived acceptability*

*Source: Freeman (2025)*

## 4.3 Gender Disparities

Gender has a significant influence on AI adoption behaviour among university students, with females being more cautious compared to their male peers. Survey findings indicate 53% of females are worried about the academic misconduct implications of AI usage, compared to 35% of males, and 51% of females are worried about AI-generated misinformation, compared to 30% of males (Freeman, 2025). Figure 3 indicates gaps in motivations, pointing to male students being more likely to use AI for skill building (36% compared to 22% among females) and experimentation (20% compared to 10%), and females being more likely to avoid risks (Freeman, 2025). These findings are consistent across greater literature on gender and technology, with females showing lower self-efficacy and greater ethical concern potentially limiting AI tool adoption (Russo et al., 2025; Koning et al., 2025). Females show more AI anxiety and a less favorable attitude, and adoption gaps are up to 14 percentage points in past education (Freeman, 2025). The gap has the potential to widen the gender digital gap, and AI literacy sectors in particular (UNESCO, 2024).

*Figure 3: Distribution of reported reasons for AI use by gender*

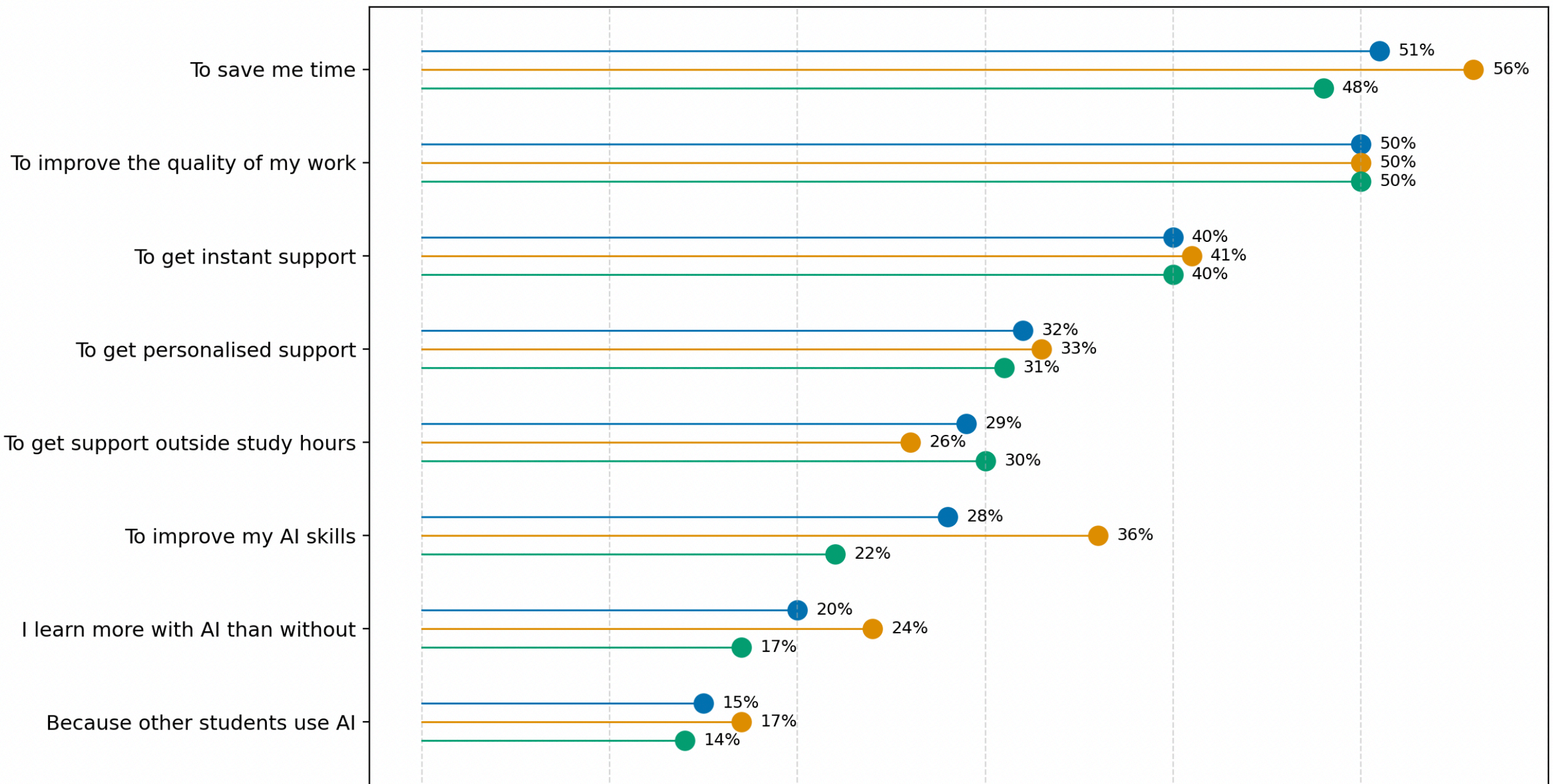

*Source: Freeman (2025)*

## 4.4 Educator and Institutional Readiness

Institutional readiness lags behind student adoption in higher education, and educators are frequently bereft of tools and training to guide AI adoption effectively. Systematic reviews mirror the finding that 72% of higher education educators have experimented with generative AI, but only 14% feel secure doing so (Cooper & Ruediger, 2025). Higher education professionals' adoption of AI has increased more than twice in the past year, 2025, but AI concerns permeate and need added support (Ellucian, 2024). Institutions offer access to generative AI tools only some 50% of the time, and fewer than 40% have comprehensive AI policies (EDUCAUSE, 2024; Brandon et al., 2025). Policy statements prioritize disclosure and ethical governance but are unevenly enforced, and institutions are thus vulnerable to ethical breaches and uneven outcomes (U.S. Department of Education, 2025; UNESCO, 2023). This readiness gap, where 79% of educators feel AI literacy was important but 45% were given training, emphasizes the importance of targeted professional development (Microsoft Corporation, 2025). Table 1 compares student and faculty metrics of readiness.

*Table 1: Comparison of AI Readiness: Students vs. Faculty*

| Group | Adoption Rate | Training Gaps | Confidence Levels |
|---|---|---|---|

| Students | 92% (Freeman, 2025) | 58% lack skills (Campus Technology, 2024) | High experimentation (86%) |
|---|---|---|---|
| Faculty | 72% experimented (Cooper & Ruediger, 2025) | 45% no training (Microsoft, 2025) | 14% confident (Cooper & Ruediger, 2025) |

*Source: Freeman (2025), Campus Technology (2024), Cooper & Ruediger (2025), and Microsoft Corporation (2025)*

## 5. Discussion

The widespread adoption of AI in higher education presents both significant challenges and transformative opportunities, as evidenced by the interplay of student motivations, ethical gaps, gender disparities, and institutional readiness (Freeman, 2025; Walter, 2024). This section synthesizes these findings, using Figures 1–3 to highlight the tensions and potential for innovation, and situates them within broader pedagogical and ethical frameworks to inform institutional strategies (Ng et al., 2024; UNESCO, 2023).

### 5.1 Challenges of AI Integration

The results indicate an AI adoption landscape in tertiary education as complicated and involves extensive student engagement, profound gaps in ethical understanding, and unprepared institutions. AI's revolutionary influence is underlined by the 92% adoption rate and 88% assessment use, but 18% of students copying AI work and only 36% receiving guidance demonstrate a profound disconnect (Freeman, 2025). Figure 2 highlights such an ethics gap with a mismatch of practices and perceived acceptability, potentially damaging academic integrity (Freeman, 2025). Demonstrating such a ``shadow pedagogy" with students creating AI workflows outside of class points to a need for institution-wide frameworks to harness enthusiasm into proper use (Walter, 2024). Gender inequalities make integration even more problematic, with greater concerns about misconduct among female students (53%) and misinformation (51%) compared to their male counterparts (35% and 30%), suggesting gender-sensitive approaches are needed (Freeman, 2025; Figure 3). Low teacher preparedness exacerbates these woes as institutions are unable to offer guidance or rethink curricula (Zawacki-Richter et al., 2024; Microsoft Corporation, 2025). These are characteristic of the more widespread concerns about the potential of algorithmic bias and data privacy to entrench inequalities unless remedied (Zheng et al., 2025; Wargo & Anderson, 2024). Without correction, the headlong rush of adoption threatens to solidify an adversarial educational landscape where ethical parameters are uncertain.

### 5.2 Opportunities for Innovation

In the face of these obstacles, the universal adoption of AI presents substantial potential for pedagogical innovation in the context of higher education. Students' quality improvement and time-saving motivations of 51% and 50%, respectively, indicate a willingness to utilize AI in enhancing their studies, as illustrated in Figure 1 (Freeman, 2025). This potential can be utilized to create AI literacy courses teaching critical thinking, ethical understanding, and technical skills, corresponding to envisaged curricula (UNESCO, 2024; Salas-Pilco et al., 2025). For example, the use of AI in assessment design, such as AI-boosted brainstorming and final analysis by humans, could create genuine learning experiences and ensure academic integrity (Ng et al., 2024; Chen et al., 2024). Gender gaps provide an opportunity to create inclusive curricula using female students' ethical sensitivity to foster fair-minded access (Russo et al., 2025). Additionally, the rising interest among educators provides potential for scalable staff development, building on successful case-study approaches (Chen et al., 2024). By framing institution-wide policies around student realities, furthering education can turn AI from a force of disruption into an agent of customized, fair education (Holmes et al., 2019; Liang et al., 2025).

As a novel contribution, this paper proposes an Ethical AI Integration Model (Table 2), with three pillars: Literacy Training (focusing on technical and ethical skills), Gender-Inclusive Policies (addressing disparities through targeted support), and Assessment Redesign (incorporating AI-transparent tasks). This model draws on findings from surveys and reviews to provide a practical framework for institutions.

*Table 2: Ethical AI Integration Model*

| Pillar | Description |
| --- | --- |
| Literacy Training | Develop programs teaching AI evaluation, bias detection, and ethical use, integrating UNESCO frameworks (2024). |
| Gender-Inclusive Policies | Implement support for underrepresented groups, leveraging insights on anxiety (Russo et al., 2025). |
| Assessment Redesign | Create AI-transparent assignments to deter misconduct while fostering critical thinking (Ng et al., 2024). |

*Source: UNESCO (2024), Russo et al. (2025), and Ng et al. (2024)*

## 6. Recommendations

To tackle AI's multiple demands and possibilities in further and higher education, shared and strategic responses are required. Institutions, policymakers, and researchers need to work together to develop resilient frameworks to underpin ethical AI practice, optimize pedagogical results, and balance inequalities of access and preparation. This requires institutionally explicit AI practice policies, investments in teacher education and inclusive courses, and interest in longitudinal research into AI's future effects. In conjunction with evidence-informed frameworks, such as UNESCO's (2023) and Freeman's (2025), these efforts allow securing of AI as a means of fair, open, and transformative education.

### 6.1 Institutional Policies

To confront the ethical and pedagogical dilemmas of AI use, universities must create explicit, comprehensive AI policies. These need to specify acceptable use according to task type and learning goals, such as permitting AI use for ideation but mandating human synthesis (Freeman, 2025; UNESCO, 2023). Redesign of assessment is needed, including AI-transparent assignments (e.g., critical reflective essays about AI use) to prevent misconduct but encourage critical thinking (Ng et al., 2024). Universities must embed compulsory AI-literacy courses, educating students to critically assess output for bias and accuracy, as proposed by recent frameworks (UNESCO, 2024). This directly addresses the ethics gap illustrated in Figure 2 (Freeman, 2025).

### 6.2 Policy and Investment

Institutions and policymakers need to make investments in teacher professional development to close the preparedness gap, and systematic analyses have emphasized the significance of focused training (Zawacki-Richter et al., 2024). Investment ought to be focused on case-study programs designing AI literacy and pedagogical confidence among teachers (Chen et al., 2024). For gender inequalities shown in Figure 3, investments ought to go into inclusive projects, assuring fair access and engagement (Freeman, 2025; Koning et al., 2025). National frameworks, including the U.S. Department of Education (2025) and UNESCO (2023), ought to be modified to ensure consistent adoption, prioritizing openness and human moderation. Secure AI tools need investment, as proposed by EDUCAUSE (2024).

### 6.3 Future Research

Future research is required to close knowledge gaps regarding AI's long-term effects. Longitudinal research must go beyond 2025 to further track adoption trajectories and outcomes, expanding Freeman's (2025) data set. Experimental evaluations of AI literacy interventions and assessment redrafts may establish best practices (Chen et al., 2024; Zawacki-Richter et al., 2024). Cross-institutional comparative research would clarify ethical use determinants and gender gaps (Russo et al., 2025). Research on AI's ability to compensate inequities, such as at the level of individual learning, would be of much interest (Salas-Pilco et al., 2025). Future longitudinal research on faculty would be valuable to establish training effects across multiple periods of time (Ellucian, 2024).

## 7. Conclusion

The explosive adoption of generative AI in college education since 2023 has transformed teaching practice, with 92% of students embracing AI devices and 88% employing them on assessments, sparked by time savings (51%) and quality enhancements (50%) (Freeman, 2025; Figure 1). But student-led adoption has gotten ahead of institution-wide preparedness, creating a ``shadow pedagogy" of unguided AI practice (Walter, 2024). The ethical gap, signalled by 18% of students plagiarizing AI-generated material, though only 20% finding it acceptable (Freeman, 2025; Figure 2), highlights the critical need for explicit guidelines. Gender inequities, with females reporting greater concern about misconduct (53%) and misinformation (51%) than males (Freeman, 2025; Figure 3), have the potential to exacerbate the digital divide. Inadequate educator preparation further compounds these difficulties (Zawacki-Richter et al., 2024; Microsoft Corporation, 2025). But potential abounds to use AI to support personalized learning and inclusive education via extensive literacy programs and assessment redesign (Ng et al., 2024; Chen et al., 2024). Higher education institutions need to move quickly to install ethical policies, provide investments in educator training, and redress gender inequities so AI bolsters rather than sabotages academic integrity and fairness (UNESCO, 2023; U.S. Department of Education, 2025). The proposed Ethical AI Integration Model provides a future-oriented model to guide this evolution, transmuting AI potential into a spark of fair, innovative education.